\providecommand{\LyX}{L\kern-.1667em\lower.25em\hbox{Y}\kern-.125emX\@}
\newcommand{\noun}[1]{\textsc{#1}}

\documentclass[acus]{JAC2000}   
\usepackage{graphics} 

\makeatother

\begin{document}

\title{Analysis of Longitudinal Bunching in an FEL Driven Two-Beam Accelerator\thanks{
The work at LBNL was performed under the auspices of the U.S. Department of
Energy by University of California Lawrence Berkeley National Laboratory under
contract No. AC03-76SF00098.
}}

\author{S. Lidia, LBNL, Berkeley, CA USA\\
J. Gardelle, T. Lefevre, CEA/CESTA, Le Barp, France\\
J.T. Donohue, CENBG, Gradignan, France \\
P. Gouard, J.L. Rullier, CEA/DIF, Bruyeres le Chatel, France\\
C. Vermare, LANL, Los Alamos, NM USA}

\maketitle
\begin{abstract}
Recent experiments {[}1{]} have explored the use of a free-electron laser (FEL)
as a buncher for a microwave two-beam accelerator, and the subsequent driving
of a standing-wave rf output cavity. Here we present a deeper analysis of the
longitudinal dynamics of the electron bunches as they are transported from the
end of the FEL and through the output cavity. In particular, we examine the
effect of the transport region and cavity aperture to filter the bunched portion
of the beam.
\end{abstract}

\section{INTRODUCTION}

Since 1995, free-electron laser (FEL) experiments at the CEA/CESTA facility
have addressed the problem of the generation of a suitable bunched drive beam
for a two-beam accelerator using linear induction accelerator technology. In
these trials, a 32 period long bifilar-helix wiggler is coupled with a 35\-GHz,
5\-kW magnetron to provide an effective FEL interaction with the beam. Early
experiments {[}2{]}, {[}3{]} demonstrated optical diagnostic techniques to show
bunching of the beam at the 35\-GHz FEL resonant frequency.

In the first cavity experiments {[}1{]}, the induction linac 'PIVAIR' was utilized,
since its design energy of 7.2\-MeV is near optimum for a Ka-band two-beam
accelerator based upon the relativistic-klystron mechanism {[}4{]}. During operation,
PIVAIR delivered a 6.7\-MeV, 3\-kA, 60\-ns (FWHM) electron beam. The emittance
out of the injector is approximately 1000\( \pi  \)~\-~mm~\-~mrad, and the
energy spread is less than 1\% (rms) over the pulse length. The full current
is collimated to 830~\( \pm  \)~30\-A at the FEL entrance. Two 6-period adiabatic
sections are used to inject the beam into the proper helical trajectory inside
the wiggler, and then to release the beam back into the transport line afterwards.

\begin{figure}
{\par\centering \resizebox*{1\columnwidth}{!}{\rotatebox{-90}{\includegraphics{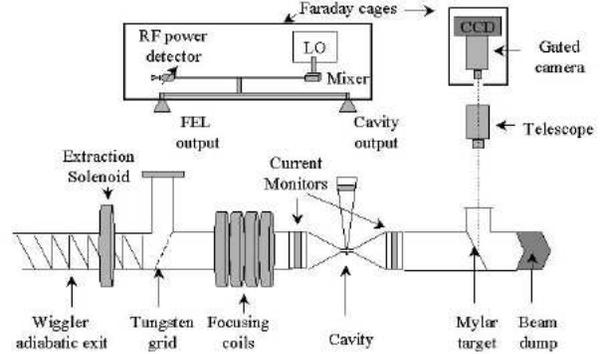}}} \par}

\caption{\label{Cavity Experiment Schematic}Schematic of downstream transport beamline,
and rf diagnostics.}
\end{figure}

After the wiggler follows a short transport beamline to capture and focus the
beam through a narrow-aperture (4\-mm ID), 35\-GHz, single-cell rf output
cavity {[}5{]}. The beamline consists of a section of stainless steel pipe (39\-mm
ID, 1.2\-m long) with a set of solenoid magnets to provide focusing through
the rf cavity. The rf power generated by the beam in the cavity is collected
and analyzed, while the beam itself is dumped. This set-up is shown in Figure
\ref{Cavity Experiment Schematic}.

\section{SIMULATION CODES}

\begin{figure}
{\par\centering \resizebox*{1\columnwidth}{!}{\rotatebox{-90}{\includegraphics{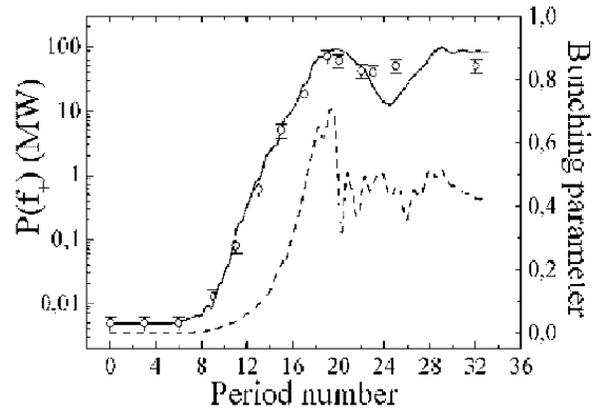}}} \par}

\caption{\label{FEL Output}FEL power and bunching evolution in the wiggler, measured
versus period number.}
\end{figure}

Two separate numerical simulation codes are used to model the system behavior.
The first is the steady-state, 3-D FEL code, \noun{}\emph{\noun{solitude}}
\noun{{[}6{]}}. The evolution of the FEL mode power, both as measured and
as calculated by \noun{solitude}, is shown (circles and solid line, respectively)
in Figure \ref{FEL Output}. Also shown is the calculated value of the bunching
parameter (dashed line) {[}7{]}.
 Not shown is the evolution of the beam current during transport through the
wiggler. Experimentally, the current exiting the wiggler was observed to be
\( \sim \,  \)250~A. This value was reproduced in the FEL simulations {[}8{]}.

\begin{figure}
{\par\centering \resizebox*{1\columnwidth}{!}{\includegraphics{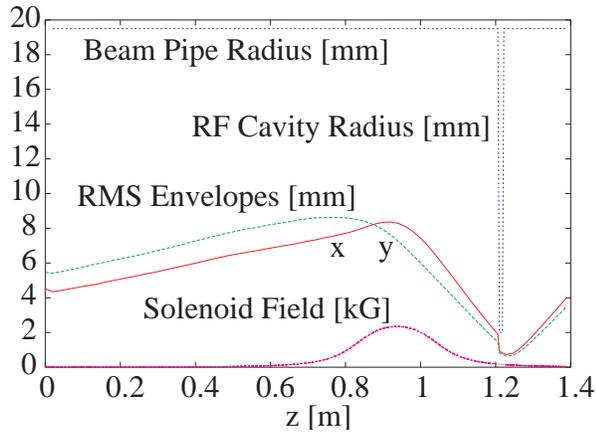}} \par}

\caption{\label{RMS Envelopes}Simulation results of beam transport from the wiggler
exit through the rf cavity.}
\end{figure}

\begin{figure}
{\par\centering \resizebox*{1\columnwidth}{!}{\includegraphics{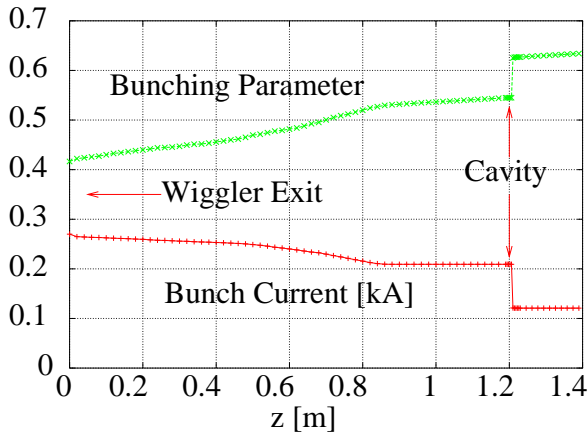}} \par}

\caption{\label{Current Transport}Simulation results of current and bunching parameter
transport.}
\end{figure}

The \noun{rks} code {[}9{]} is then used to propagate the beam from the end
of the wiggler through the cavity, and to calculate the interaction of the beam
with the rf output structure. The 6\-D particle distribution of the beam at
the end of the wiggler, as calculated by \noun{solitude}, is used as input
to \noun{rks}. The evolution of the beam rms envelopes are shown in Figure
\ref{RMS Envelopes}.
 The cavity acts as a collimator, reducing the beam current, as can be seen
in Figure \ref{Current Transport}.
 This degree of current loss was observed experimentally. The calculated power
developed in the rf cavity is also comparable to that observed experimentally
{[}1{]}.

\section{FILTERING AND BUNCH ENHANCEMENT}

The interesting feature to observe in Figure \ref{Current Transport}, is the
discontinuous growth of the bunching parameter, and simultaneous current loss,
as the beam is partially focused through the cavity. This is preceded by the
gradual loss of current, and the gradual increase in the bunching parameter
in the transport region between the end of the wiggler and the entrance to the
rf cavity. As was pointed out in {[}1{]}, the transport line and cavity appear
to act as a filter that preferentially selects the bunched portion of the beam
for transmission. We seek to analyze this behavior in terms of the dynamics
of the beam in the transport line following the wiggler. 

\begin{figure}
{\par\centering \resizebox*{1\columnwidth}{!}{\includegraphics{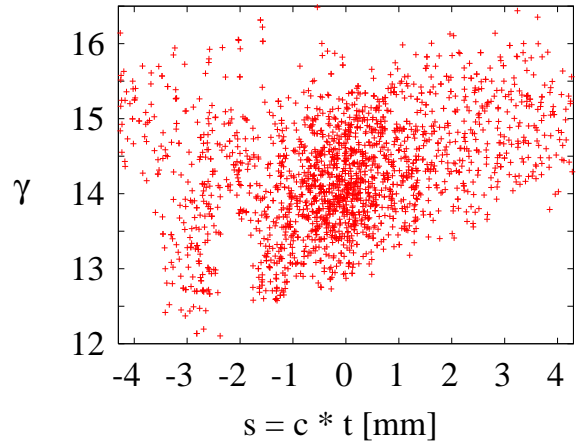}} \par}

\caption{\label{Initial phase space}Longitudinal phase space distribution at wiggler
exit.}
\end{figure}

The longitudinal phase space of the bunches near the middle of the beam pulse
at the exit of the wiggler are shown in Figure \ref{Initial phase space}. As
shown, there is significant initial bunching (b \( \sim  \) 0.4) as well as
'tilt' (energy-phase correlation). The presence of this tilt arises from the
fortuitous extraction of the beam at an appropriate synchrotron oscillation
phase in the 'saturated' regime of the FEL interaction. This tilt contributes
to continued bunching in a ballistic transport line. The effect of space charge
forces upon debunching are limited by the low current (250 A) and high kinetic
energy (6.7 MeV), and significant debunching will only appear after several
meters {[}10{]}. This tilt can account for a modest rise in the bunching parameter,
from 0.4 to \( \sim  \)0.5, as discussed below.

\begin{figure}
{\par\centering \resizebox*{1\columnwidth}{!}{\includegraphics{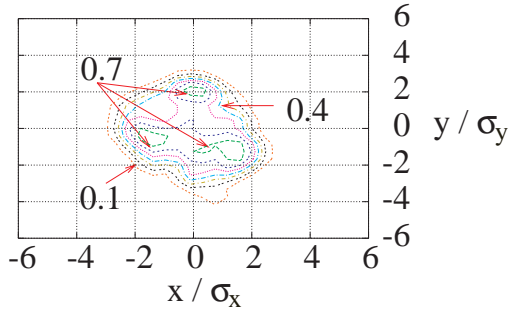}} \par}

\caption{\label{Initial bunching}Transverse distribution of bunching parameter at wiggler
exit.}
\end{figure}

In addition to the energy tilt, the bunches emerging from the wiggler exhibit
nonuniform bunching over the transverse distribution. This is shown in Figure
\ref{Initial bunching}.
 Displayed are contours of constant bunching parameter as a function of transverse
position. The transverse position coordinate has been normalized by the appropriate
rms transverse beam size (\( \sigma _{x} \) or \( \sigma _{y} \)). While the
average value of the bunching parameter is \( \sim  \)0.4, there is a large
degree of variation with the high-brightness, central core more strongly bunched
than the outlying edges. Collimation of the beam can then strip away the less-bunched
regions, resulting in an overall enhancement of the average bunching parameter.

The origin of the transverse variation can be related to the variation of the
electromagnetic signal co-propagating with the beam in the waveguide of the
FEL. Optical guiding studies {[}11{]} show that both the beam density and the
electromagnetic mode amplitude decrease with increasing transverse distance
from the beam axis. There is, then, a smaller coupling between the beam and
the mode at distances from the beam axis, with the subsequent decrease in the
forces responsible for bunching.

\begin{figure}
{\par\centering \resizebox*{1\columnwidth}{!}{\includegraphics{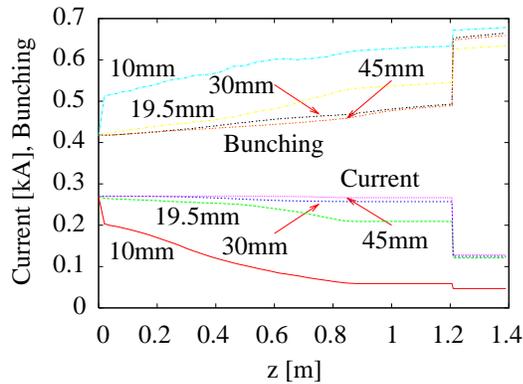}} \par}

\caption{\label{Pipe Diameter}Evolution of beam current and bunching parameter with
varying beam pipe diameters.}
\end{figure}

\begin{figure}
{\par\centering \resizebox*{1\columnwidth}{!}{\includegraphics{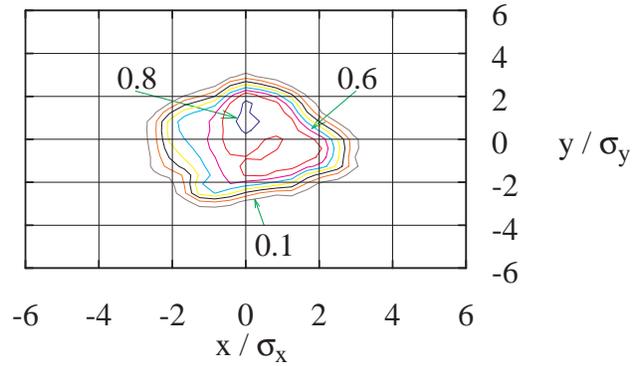}} \par}

\caption{\label{Final bunching}Transverse distribution of bunching parameter at cavity
exit.}
\end{figure}

A series of simulations were performed in which the beam pipe radius of the
transport line between the wiggler and rf cavity was varied. The purpose of
this was to explore the relative influence of the two bunching effects described
above. The results are shown in Figure \ref{Pipe Diameter}, showing evolution
of beam current and bunching parameter along the beamline.
 As shown, the smaller pipes act as collimating agents, while the larger pipes
transmit nearly 100\% of the beam current from the wiggler to the entrance of
the cavity. In the simulations, the smaller beam pipes allowed the less-bunched
portions of the beam to be intercepted, thereby increasing the average bunching
parameter. However, all simulations demonstrated ballistic bunching due to the
energy-phase tilt. At the end of the transport line lies the cavity with a 2mm
bore radius, which acts as a final collimator and limits drastically the percentage
of transmitted current, while also stripping away the unbunched portions from
the highly-bunched core. The final transverse distribution of the bunching parameter
is shown in Figure \ref{Final bunching}, taken at the exit plane of the cavity.
 This distribution exhibits a broader central plateau with a greater degree
of bunching than seen in Figure \ref{Initial bunching}.

\section{CONCLUSIONS}

We have presented results of simulations to assist in the analysis of experimental
measurements of current and bunching transport in a high-frequency, two-beam
accelerator prototype experiment. We have shown that the average degree of longitudinal
bunching in a beam that exits an FEL amplifier can be improved by collimation.
However, this may also be accompanied by significant loss of current.

\section{REFERENCES}

\noindent {[}1{]} T. Lefevre, et. al., \emph{Phys. Rev. Lett.} \textbf{84} (2000),
1188.

\noindent {[}2{]} J. Gardelle, et. al., \emph{Phys. Rev. Lett.} \textbf{76}
(1996), 4532.

\noindent {[}3{]} J. Gardelle, et. al., \emph{Phys. Rev. Lett.} \textbf{79}
(1997), 3905.

\noindent {[}4{]} G. Westenskow, et. al.\emph{, Proceedings of the VIII International
Workshop on Linear Colliders}, Frascati (1999).

\noindent {[}5{]} S.M. Lidia, et. al., \emph{Proceedings of the XIX International
Linear Accelerator Conference}, Chicago (1998), 97.

\noindent {[}6{]} J. Gardelle, et. al., \emph{Phys. Rev. E} \textbf{50} (1994),
4973.

\noindent {[}7{]} The bunching parameter is defined as \( b=\left| \left\langle \exp \left( i\psi \right) \right\rangle \right|  \),
where \( \psi =k_{w}z-\omega \left( z/c-t\right)  \) is the usual phase variable
of an electron in the bunch, and where \( \left\langle \right\rangle  \) denotes
the bunch ensemble average.

{\par\raggedright {[}8{]} S.M. Lidia, et. al., \emph{Proceedings of the 1999
IEEE Particle Accelerator Conference}, New York (1999), 1797.\par}

\noindent {[}9{]} S.M. Lidia, \emph{Proceedings of the 1999 IEEE Particle Accelerator
Conference}, New York (1999), 2870.

{\par\raggedright {[}10{]} J. Gardelle, et. al., \emph{Proceedings of the XIX
International Linear Accelerator Conference}, Chicago (1998), 794.\par}

{\par\raggedright {[}11{]} A. Bhattacharjee, et. al., \emph{Phys. Rev. Lett.}
\textbf{60} (1988), 1254.\par}

\end{document}